\documentclass[twocolumn]{aastex62}

\usepackage{natbib}
\usepackage{subfigure}
\setcitestyle{sort,comma}
\usepackage{csvsimple}

\usepackage{amsmath}
\usepackage{fancyvrb}
\usepackage{multirow}
\usepackage{tabularx}
\newcommand\setrow[1]{\gdef\rowmac{#1}#1\ignorespaces}
\newcommand\tpm{$\pm\text{ }$}
\newcommand\clearrow{\global\let\rowmac\relax}
\clearrow
\usepackage{graphicx}
\usepackage{array}
\usepackage{placeins}
\usepackage{hyperref}

\usepackage{lineno}
\shorttitle{JWST MIRI Eclipse Reanalysis using FN-PCA}
\shortauthors{}

\begin{document}

\title{\large{Uniform Reanalysis of JWST MIRI 15$\mu$m Exoplanet Eclipse Observations using\\ Frame-Normalized Principal Component Analysis}}

\correspondingauthor{Nicholas J. Connors}
\email{nicholas.connors@umontreal.ca}

\author[0000-0001-5848-6750]{Nicholas J. Connors} 
\affil{Department of Physics and Trottier Institute for Research on Exoplanets, Universit\'{e} de Montr\'{e}al, Montreal, QC, Canada}

\author[0009-0005-9152-9480]{Christopher Monaghan} 
\affil{Department of Physics and Trottier Institute for Research on Exoplanets, Universit\'{e} de Montr\'{e}al, Montreal, QC, Canada}

\author[0000-0001-5578-1498]{Bj\"{o}rn Benneke} 
\affil{Department of Earth, Planetary, and Space Sciences, University of California, Los Angeles, CA, USA}
\affil{Department of Physics and Trottier Institute for Research on Exoplanets, Universit\'{e} de Montr\'{e}al, Montreal, QC, Canada}

\author[0000-0003-4987-6591]{Lisa Dang} 
\affil{Department of Physics and Trottier Institute for Research on Exoplanets, Universit\'{e} de Montr\'{e}al, Montreal, QC, Canada}





\begin{abstract}

JWST MIRI $15\,\micron$ time-series eclipse photometry presents a powerful way to probe for the presence of atmospheres on low-temperature rocky exoplanets orbiting nearby stars. Here, we introduce a novel technique, frame-normalized principal component analysis (FN-PCA) to analyze and detrend these MIRI time-series observations. Using the FN-PCA technique, we perform a uniform reanalysis of the published MIRI $15\micron$ observations of LHS\,1478\,b, TOI-1468\,b, LHS\,1140\,c, TRAPPIST-1\,b, and TRAPPIST-1\,c using our new data reduction pipeline (\verb!Erebus!) and compare them to different potential atmospheric and surface compositions. We also investigate additional public data sets with the sole purpose of understanding the instrument systematics affecting MIRI. We identify and categorize important detector-level systematics in the observations that are generally present across all 17 analyzed eclipse observations, which we illustrate as eigenimage/eigenvalue pairs in the FN-PCA. One of these eigenimage/eigenvalue pairs corresponds to the prominent ramp effect at the beginning of the time-series observations which has widely been reported for JWST and Spitzer photometry. For JWST/MIRI, we show that the detector settling time scales exponentially with the apparent magnitude of the target star $T_\mathrm{set}\,\mathrm{[hours]} = 0.063\exp^{0.427\cdot m_K} -0.657$. This uniform reanalysis and investigation of JWST/MIRI systematics is done in preparation for the 500 hour Rocky Worlds DDT survey, to demonstrate a data-driven systematic model usable across all MIRI $15\,\micron$ datasets.

\end{abstract}



\section{Introduction} \label{sec:intro}

Among the known rocky exoplanets, those around M-dwarfs are considered the best small targets for atmospheric characterization. The low temperatures of these stars allow habitable zone exoplanets to orbit with short periods, giving frequent opportunities to observe temperate planets in eclipse \citep{ddt_working_group, m_dwarf_hab_zone}. Additionally, low-mass dwarf-type stars are the most common star type in our galaxy, giving an abundance of targets for characterization \citep{m_dwarf_frequency}. However, the stellar activity of these stars may make it difficult for planets orbiting them to form or sustain atmospheres \citep{m_dwarf_atmo_loss}. Understanding if this population of planets can have atmospheres is therefore an important step in the search for habitable worlds similar to our own \citep{ddt_working_group, 2024_roadmap}.

Secondary eclipse photometry using the MIRI instrument at 15 $\mu$m on board the James Webb Space Telescope allows us to infer the atmospheric and surface composition of rocky exoplanets  without being subject to stellar contamination found in exoplanet transit observations \citep{rackham_2018_tls_effect, zieba_no_2023, greene_thermal_2023, august_hot_2024, hotrocks2, fortune2025hotrockssurveyiii}. Despite the effects of tidal locking, atmospheric models predict that it is still possible for these short-period planets orbiting M dwarfs to maintain atmospheres \citep{turbet2018}. The presence of an atmosphere allows for heat-redistribution from the day-side to the night-side, resulting in a shallower eclipse depth compared to the bare rock scenario \citep{Koll_2022, coy2025populationlevelhypothesistestingrocky}.

Measurements of secondary eclipses require high precision to distinguish between the different atmospheric or surface composition models for a given planet. For secondary eclipse observations performed using MIRI F1500W it is important to understand the systematic noise of the instrument to improve precision \citep{zieba_no_2023, greene_thermal_2023, august_hot_2024, hotrocks2, fortune2025hotrockssurveyiii}. 

So far there have been two completed JWST programs performing MIRI 15 $\mu$m secondary eclipse photometry to infer the atmospheric compositions of rocky exoplanets orbiting M-dwarf stars: TRAPPIST-1\,b \citep[GTO-1177][]{greene_thermal_2023} and TRAPPIST-1\,c \citep[GO-2304][]{zieba_no_2023}. There has also been a full phase curve observation of TRAPPIST-1b and c using the same instrument (GO-3077). There are currently two more secondary-eclipse programs in-progress: The Hot Rocks Survey observing 9 exoplanets \citep[GO-3730][]{august_hot_2024, hotrocks2} and a program for observing the potentially volcanically active rocky exoplanet LP\,791-18\,d \citep[GO-6457][]{2024jwst.prop.6457B}. Future observations include a recently approved Cycle 4 program to follow up on previous observations of LHS\,1478\,b using MIRI LRS and F1500W \citep[GO-7675][]{2025jwst.prop.7675A}, and the 500 hour Rocky Worlds Director's Discretionary Time survey which will investigate a number of rocky exoplanets in combination with the Hubble Space Telescope \citep{ddt_working_group}. The targets LTT\,1445\,A\,c, LTT\,1445\,A\,b, LHS\,1140\,b, and GJ\,3929\,b are already announced for this survey.

The currently published MIRI 15$\mu$m secondary eclipse literature uses a variety of methods to detrend against systematic noise. This includes using polynomials from the 0th to 3rd degree of time, polynomials of the Gaussian centroid positions and widths, exponential ramps in time, Gaussian processes, and linear combinations of up to 36 pixel time-series, as well as different combinations of each \citep{zieba_no_2023, greene_thermal_2023, august_hot_2024, hotrocks2}. The final choice of which combination of models to use is then justified by goodness of fit metrics, and all models are parametric and make assumptions on the behaviour of the instrument noise.

In this paper we present \verb!Erebus!, a new data analysis pipeline for JWST/MIRI photometric data of transiting exoplanets, and use it to analyze F1500W data. It uses a novel data-driven method of detrending lightcurves against systematic noise: Frame-Normalized Principal Component Analysis (FN-PCA). We perform principal component analysis on the time-series image observations of a star during a secondary eclipse, after normalizing by the total intensity of each frame. This removes the effects of astrophysical signals (e.g., the eclipse itself and stellar variability) from the eigenvalues generated by the PCA, as we expect such signals to show as uniform increases/decreases in intensity across the entire detector. We detrend the lightcurve against the time-series eigenvalues of the top 5 principal components found for each dataset, which represent pixel-level systematics of the detector during the observation. While we only use it for F1500W data in this paper, \verb!Erebus! is generally applicable to the other MIRI filters, notably the F1280W filter which has also been used for secondary eclipse observations \citep{Ducrot_2024}.

PCA is a data-analysis tool which reduces the dimensionality of a data set by re-expressing it as a linear combination of eigenvalue/eigenvector pairs called principal components. These principal components are ranked in order of variance, so that those with a lower variance are classified as noise and those with a higher variance are classified as important signals. The principal components are selected such that they are all orthogonal to each other. This technique is used to find underlying structures in otherwise confusing datasets \citep{shlens2014tutorialprincipalcomponentanalysis}.

Past analyses of JWST exoplanet time series observations have shown that small changes in PSF morphology can be precisely tracked via principal component analysis of the detector images, proving useful in detrending against distinct sources of instrumental noise \citep{ahrer_2025ApJ, pca_lp1, pca_lp2, radica_2024ApJ, radica_2025ApJ}.

We use this new method to reanalyze existing secondary eclipse observations performed with MIRI Imaging F1500W for the planets  LHS\,1478\,b \citep{august_hot_2024}, TOI-1468\,b \citep{hotrocks2}, LHS\,1140\,c \citep{fortune2025hotrockssurveyiii}, TRAPPIST-1\,b \citep{greene_thermal_2023}, and TRAPPIST-1\,c \citep{zieba_no_2023}. 

This paper is divided into 5 sections. Section \ref{sec:data} will describe the methodology used for extracting secondary eclipse lightcurves which are then detrended and analyzed using FN-PCA. Section \ref{sec:systematics} will show how the principal components found using FN-PCA give us insights into telescope and detector systematics. Section \ref{sec:scientific-results} will compare our results to different potential atmospheric and surface compositions of the five planets we reanalyzed. Section \ref{sec:conclusion} summarizes our conclusions.

\section{Data Analysis}\label{sec:data}

\begin{deluxetable*}{c c c c c c c c}[tb]
\def\arraystretch{1.2}
\setlength{\tabcolsep}{1pt} 
\tabletypesize{\scriptsize}
\tablecaption{Orbital parameters used to determine Gaussian priors. \label{table:inputs}}
\tablehead{
\colhead{Planet} & \colhead{$t_0$ (BJD-2,450,000)} & \colhead{$P$ (days)} & \colhead{$R_p/R_*$} & \colhead{$a/R_*$} & \colhead{$i$ (deg)} & \colhead{$e$} & \colhead{$\omega$ (deg)}
}
\startdata
TRAPPIST-1\,b 
& \textbf{*}
& $	1.51088432(15)$ $^{(1)}$
& $0.08590(37)$ $^{(2)}$
& $20.843^{+0.094}_{-0.155}$ $^{(2)}$
& $89.728(165)$ $^{(2)}$
& $0.00622(304)$ $^{(7)}$
& $336.86\pm34.24$ $^{(7)}$
\\
TRAPPIST-1\,c
& \textbf{*}
& $2.42179346(23)$ $^{(1)}$ 
& $0.08440(38)$ $^{(2)}$
& $28.549^{+0.129}_{-0.212}$ $^{(2)}$
& $89.778(118)$ $^{(2)}$
& $0.00654(188)$ $^{(7)}$
& $282.45\pm17.10$ $^{(7)}$
\\
TOI-1468\,b
& $8765.68079^{+0.00070}_{-0.00069}$ $^{(3)}$ 
& $1.8805136^{+0.0000024}_{-0.0000026}$ $^{(3)}$
& $0.03411(15)$ $^{(3)}$
& $13.14^{+0.21}_{-0.24}$ $^{(3)}$
& $88.47^{+0.34}_{-0.29}$ $^{(3)}$
& -
& -
\\
LHS\,1478\,b 
& $8786.75425(42)$ $^{(4)}$
& $1.9495378^{+0.0000040}_{-0.0000041}$ $^{(4)}$
& $0.0462^{+0.0011}_{-0.0010}$ $^{(4)}$
& $16.119^{+0.088}_{-0.094}$ $^{(4)}$
& $87.452^{+0.052}_{-0.048}$ $^{(4)}$
& -
& -
\\
LHS\,1140\,c
& $8389.2939(2)$ $^{(5)}$
& $3.777940(2)$ $^{(5)}$
& $0.05486(13)$ $^{(6)}$
& $26.57(5)$ $^{(6)}$
& $89.80(19)$ $^{(5)}$
& -
& -
\enddata
\setlength{\tabcolsep}{6pt} 
\tabletypesize{\normalsize}
\tablecomments{The larger value was used as standard deviation for the prior when the reported uncertainties are asymmetric.
\\\textbf{*:} To account for transit-timing variation, we use the closest forecasted transit time calculated in \cite{agol2024updatedforecasttrappist1times} for each individual visit of TRAPPIST-1\,b and c.
\\\textbf{References:} 
(1) {\cite{ducrot_trappist-1_2020}},
(2) {\cite{agol_2021}},
(3) {\cite{Chaturvedi}},
(4) {\cite{soto_2021}},
(5) {\cite{cadieux2023newmassradiusconstraints}},
(6) {\cite{Kokori_2023}},
(7) {\cite{grimm_2018}}}
\setlength{\tabcolsep}{6pt} 
\tabletypesize{\normalsize}
\end{deluxetable*}

\begin{figure*}[htbp]
\centering
\includegraphics[width=0.95\linewidth]{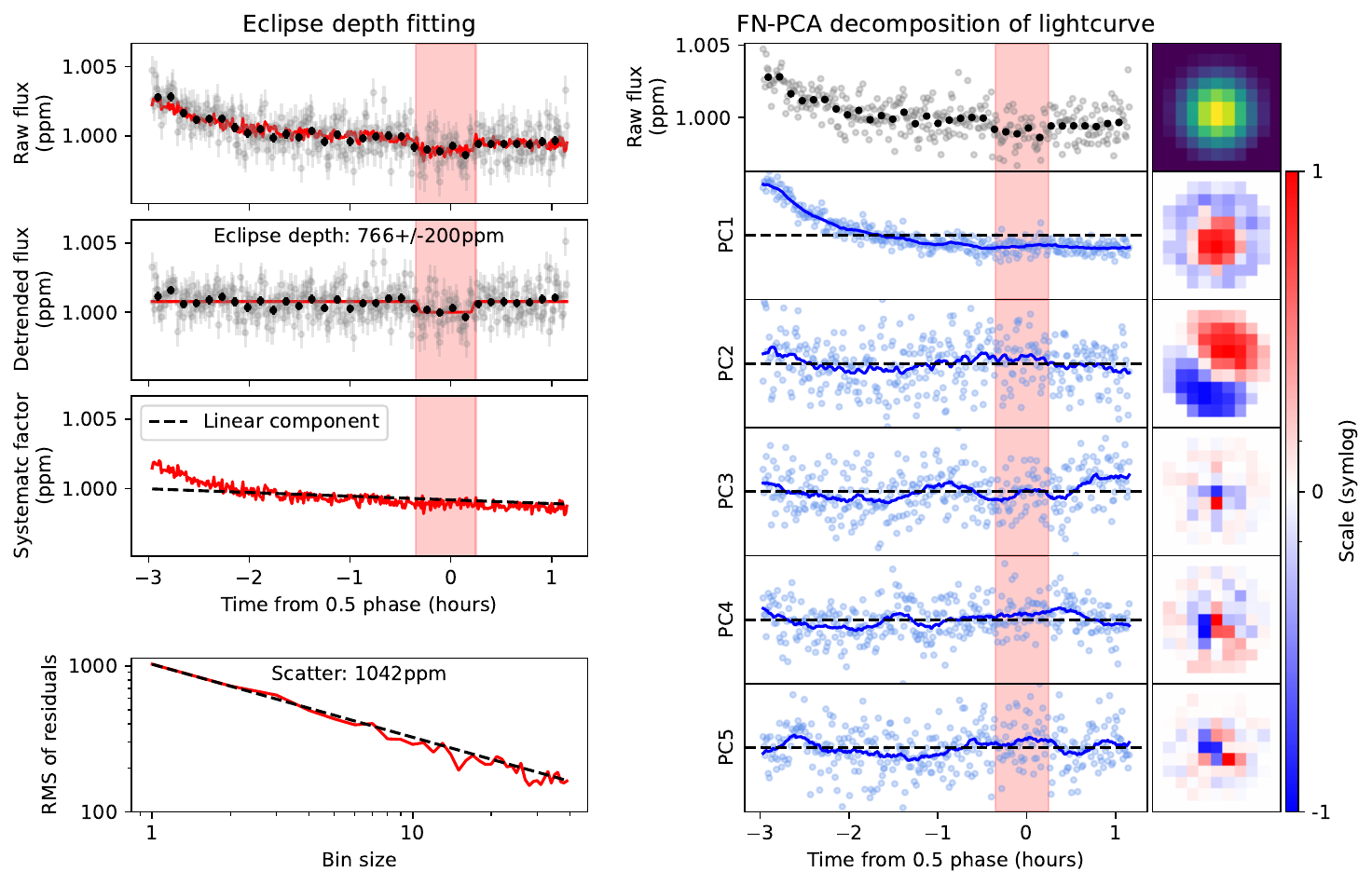}
\caption{Frame normalized principal component analysis (FN-PCA) detrending fit for TRAPPIST-1\,b visit 1 using the \texttt{Erebus} pipeline. The panels on the left, from top to bottom, show the raw lightcurve produced by our pipeline, the detrended light-curve using the FN-PCA systematic model, a breakdown of the linear and PCA components of the systematic model, and a plot comparing the root mean squared of residuals depending on bin size showing the goodness of fit compared to the ideal case (dotted-line). The panels on the right show the raw lightcurve and point spread function (first row), and the eigenvalue and eigenimages for the top 5 principal components explaining most of the variance (rows 2 to 6). The dashed line on the eigenvalue plots shows where the value is 0. As the dimmest star analyzed, we see the detector-settling component (\ref{sec:detector-settling-res}) is the most prominent principal component (PC1), followed by the positive-diagonal centroiding component (PC2) (\ref{sec:centroiding-res}). The negative-diagonal centroiding component is not present, and the remaining principal components are pixel-dominated (PC3) (\ref{sec:pixel-dominated-res}) or random (PC 4 and 5) (\ref{sec-random-res}). The predicted eclipse is shown as the red area centered around 0.5 phase.}
\label{fig:full plot}
\end{figure*}

When extracting the photometric data we use the \verb!jwst! pipeline \citep{jwst_pipeline} stage 2 outputs (\verb!_calints.fits!) available on the Mikulski Archive for Space Telescopes (MAST). For a single visit of TRAPPIST-1b (visit 5) we find the default stage 2 \verb!jwst! output on MAST to be unusable, and instead use the default stage 2 output of the \verb!Eureka!! pipeline \citep{Bell2022} ran on the uncalibrated fits data from MAST. We measure the flux within a five pixel radius aperture, and use an annulus spanning 12 to 20 pixels to perform background subtraction without overlapping the airy ring. We choose to use a uniform five pixel aperture across all data sets based on the optimal aperture sizes found in previous literature \citep{greene_thermal_2023, august_hot_2024, hotrocks2, fortune2025hotrockssurveyiii} and additionally try a four pixel aperture closer to the size used in \cite{zieba_no_2023} but find the difference in eclipse depths between the two sizes to be negligible. We mask out values marked \verb!DO NOT USE! and 3$\sigma$ outliers and replace them with 2d linearly interpolated data within the affected frame. From this we get the raw lightcurve (shown in the top panel of Figure \ref{fig:full plot}) using aperture photometry which we later detrend with our FN-PCA approach.

We perform the FN-PCA of the background-subtracted time-series image data using \verb!PCA! from the \verb!sklearn! Python library \citep{scikit-learn} with each frame normalized in order to avoid fitting out the eclipse itself. For each principal component we get a time-series of eigenvalues and an eigenimage (Figure \ref{fig:full plot}, right). The first five principal components (ranked in order of significance i.e., how much of the variance they explain) are used to detrend the lightcurve. We fit five coefficients (one per principal component) with uninformative uniform priors (from $-10$ to $10$) to create our systematic model which linearly detrends the data against the eigenvalues expressed in parts-per-million. This systematic model is multiplied by a linear slope to account for changes in stellar brightness over the observation. The principal component detrending is intended to model the systematic noise inherent in the instrument, while the linear component is to account for long-term astrophysical trends.

The systematic model is:

\begin{equation}
    F_{sys} = (1 + \sum_{i=1}^5 c_i \lambda_i) (a t + b)
\end{equation}

\noindent where $c_i$ is the coefficient for the $i$th eigenvalue $\lambda_i$, and $a$ and $b$ are the coefficients for a 1st degree polynomial.

When fitting for the light curve we use a \verb!batman! model \citep{batman} multiplied by our systematic model. We use Gaussian priors on the mid-transit time $t_0$, orbital period length $P$, planet radius $R_p$, semi-major axis $a$,  inclination $i$, as well as the eccentricity and longitude of periastron with $e\cos\omega$ and $e\sin\omega$ based on the error-bars of the latest values found on the exoplanet archive (see Table \ref{table:inputs}). We use an uninformative uniform prior for the eclipse depth from $-2000$ to $2000$ ppm. This prior allows for negative eclipse depth values to not artificially bias the fitting towards eclipse detections. We use $e\cos\omega$ to fit for the eclipse time offset from a perfectly circular orbit as described in \cite{Charbonneau_2005}:

\begin{equation}
    \Delta t_{eclipse} \approx \frac{2P}{\pi} e \cos \omega.
\end{equation}

We perform our fitting using a Markov chain Monte Carlo (MCMC) using the \verb!emcee! Python library \citep{Foreman_Mackey_2013}. For each visit, we check for convergence by running two chains in parallel and testing the Gelman-Rubin statistic \citep{gelman-rubin} to see that they have converged to the same value, and we test that the length of the chains is at least 50 times greater than the autocorrelation times of all parameters.

We perform individual fits for each visit and a joint fit for each planet. For the joint fits we bin the data from each visit by four, and we fix most orbital parameters to their values in Table \ref{table:inputs} fitting only for shared values of $e\cos\omega$ and $e\sin\omega$ to fit for the eclipse timing offset (unless assuming a circular orbit), $f_p$ for the shared eclipse depth, and individual systematic parameters for each visit. 

\begin{figure*}[htb]
    \centering
    \includegraphics[width=\linewidth]{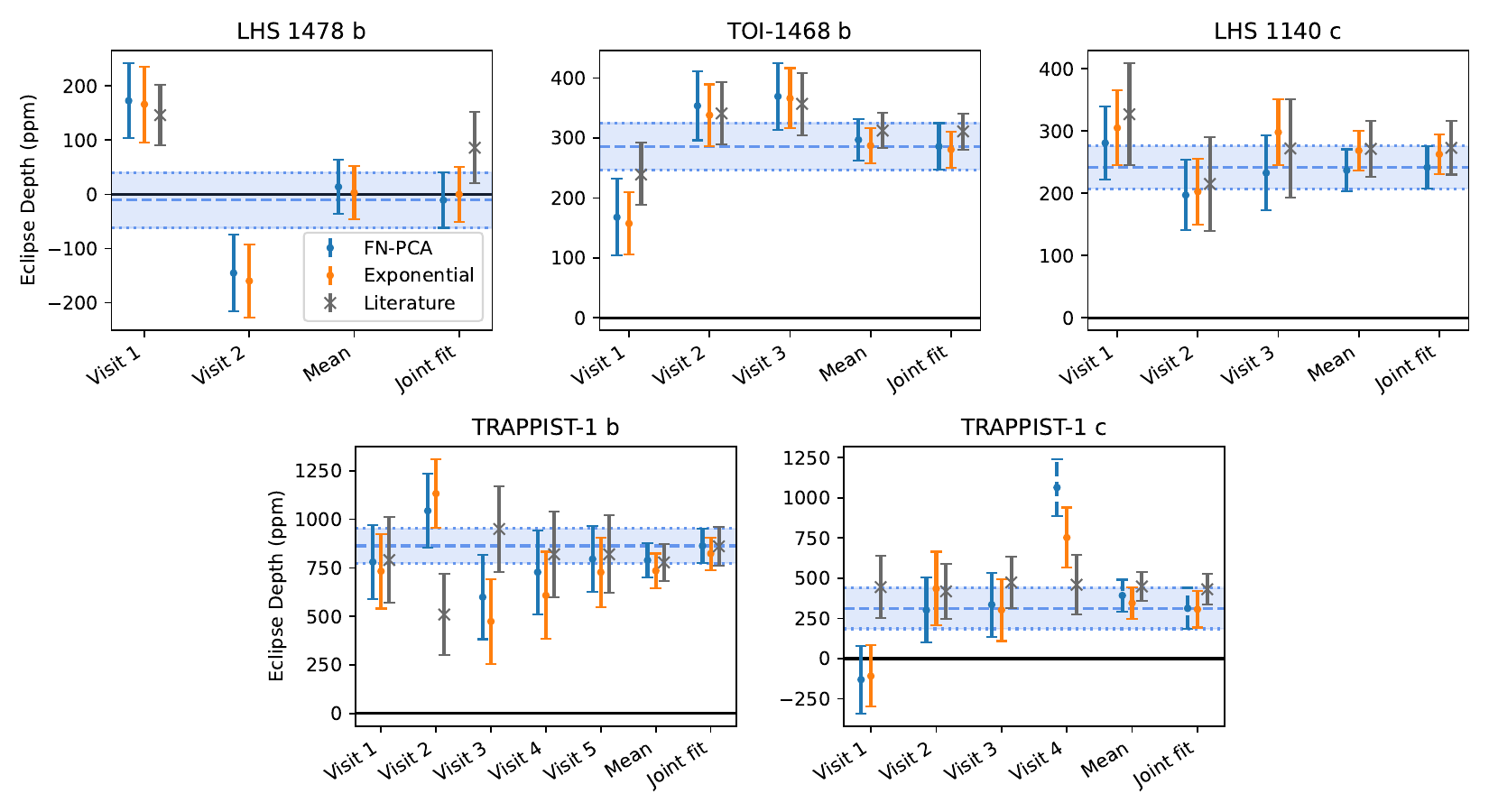}
    \caption{Eclipse depths from our reanalysis using the \texttt{Erebus} pipeline with FN-PCA and exponential fit detrending, compared to literature values. The joint fit FN-PCA result is shown by the blue dashed line with $1\sigma$ confidence interval. Exact values used to make this plot are shown in Table \ref{table:depths}. The FN-PCA analysis of TRAPPIST-1\,c visit 4 (dashed error bar) uses a 2nd degree polynomial to model the stellar variation during the visit.}
    \label{fig:eclipse_depths}
\end{figure*}

We ran our eclipse depth pipeline on LHS\,1478\,b, TOI-1468\,b, LHS\,1140\,c, TRAPPIST-1\,b, and TRAPPIST-1\,c data. For TRAPPIST-1\,b and c we fit for the eccentricity of their orbits. For the other planets we model them as entirely circular as the uncertainty in their mid-transit times and periods allow for a wider range of possible eclipse times. We discard the first and last 10 integrations of each visit. For LHS\,1478\,b we discard the last 40, as both of its visits have large systematic effects at the end of the observation as seen in \cite{august_hot_2024}. We detrend the data against the first five principal components found by FN-PCA (Figure \ref{fig:full plot}), and then against an exponential ramp for comparison. We compare these results to the results in the literature in Figure \ref{fig:eclipse_depths}, with exact values in Table \ref{table:depths}.

\begin{deluxetable*}{c c c c c}[ht]
\def\arraystretch{0.95}
\setlength\tabcolsep{6pt}
\tablecaption{Eclipse depths from this work compared to literature values.\label{table:depths}}
\tablehead{
\colhead{Planet} & \colhead{Visit \#} & \colhead{FN-PCA (ppm)} & \colhead{Exponential (ppm)} & \colhead{Literature (ppm)}
}
\startdata
\multirow{4}{*}{\textbf{LHS\,1478\,b}}
& 1 & 173 \tpm 69 & 166 \tpm 70 & 146 \tpm 56 \\
& 2 & -145 \tpm 71 & -160 \tpm 68 & - \\
\cline{2-5}
\setrow{\bfseries} & Mean & 14 \tpm 50 & 3 \tpm 49 & 146 \tpm 56 \\
\setrow{\bfseries} & Joint Fit & -11 \tpm 51 & -0 \tpm 50 & 86 \tpm 66 \\\hline
\multirow{5}{*}{\textbf{TOI-1468\,b}}
& 1 & 168 \tpm 64 & 157 \tpm 52 & $239^{+50}_{-53}$ \\
& 2 & 354 \tpm 58 & 338 \tpm 51 & $341^{+52}_{-53}$ \\
& 3 & 369 \tpm 55 & 366 \tpm 50 & $357^{+52}_{-52}$ \\
\cline{2-5}
\setrow{\bfseries} & Mean & 297 \tpm 34 & 287 \tpm 30 & $312^{+30}_{-30}$ \\
\setrow{\bfseries} & Joint Fit & 286 \tpm 39 & 281 \tpm 30 & $311^{+31}_{-30}$ \\\hline
\multirow{5}{*}{\textbf{LHS\,1140\,c}}
& 1 & 281 \tpm 59 & 305 \tpm 60 & 327 \tpm 82 \\
& 2 & 197 \tpm 57 & 203 \tpm 53 & 215 \tpm 75 \\
& 3 & 233 \tpm 60 & 298 \tpm 53 & 272 \tpm 79 \\
\cline{2-5}
\setrow{\bfseries} & Mean & 237 \tpm 34 & 269 \tpm 32 & 271 \tpm 45 \\
\setrow{\bfseries} & Joint Fit & 242 \tpm 35 & 263 \tpm 32 & 273 \tpm 43 \\\hline
\multirow{7}{*}{\textbf{TRAPPIST-1\,b}}
& 1 & 780 \tpm 191 & 733 \tpm 192 & 790 \tpm 220 \\
& 2 & 1044 \tpm 192 & 1132 \tpm 178 & 510 \tpm 210 \\
& 3 & 599 \tpm 217 & 474 \tpm 219 & 950 \tpm 220 \\
& 4 & 728 \tpm 216 & 608 \tpm 225 & 820 \tpm 220 \\
& 5 & 795 \tpm 170 & 727 \tpm 180 & 820 \tpm 200 \\
\cline{2-5}
\setrow{\bfseries} & Mean & 789 \tpm 88 & 735 \tpm 89 & 778 \tpm 96 \\
\setrow{\bfseries} & Joint Fit & 863 \tpm 90 & 823 \tpm 83 & 861 \tpm 99 \\\hline
\multirow{6}{*}{\textbf{TRAPPIST-1\,c}}
& 1 & -131 \tpm 211 & -109 \tpm 190 & 445 \tpm 193 \\
& 2 & 301 \tpm 202 & 435 \tpm 230 & 418 \tpm 173 \\
& 3 & 335 \tpm 199 & 302 \tpm 193 & 474 \tpm 158 \\
& 4 & 1064 \tpm 177 & 753 \tpm 187 & 459 \tpm 185 \\
\cline{2-5}
\setrow{\bfseries} & Mean & 393 \tpm 98 & 345 \tpm 100 & 449 \tpm 89 \\
\setrow{\bfseries} & Joint Fit & 312 \tpm 128 & 306 \tpm 113 & $431^{+97}_{-96}$ \\\hline
\enddata
\tablecomments{When a paper presents multiple results from different data reduction methods we specify the exact result we are referring to in the literature. TRAPPIST-1\,b literature results are from \cite{greene_thermal_2023}, the joint-fit result is their adopted value and the individual fits are from ``individual fits no. 1". TRAPPIST-1\,c literature results are from \cite{zieba_no_2023}, the individual fit results are from data analysis ``ED", and the joint fit result is from data analysis ``SZ". TOI-1468\,b literature results are from \cite{hotrocks2}.  LHS\,1478\,b literature results are from \cite{august_hot_2024}, the joint-fit result is from section 4.3 ``Joint fit with \texttt{juliet}", and no result is given for visit 2 (only an upper bound of 300ppm) due to systematic noise. LHS\,1140\,c literature results are from the ``GP aperture extraction" fits in \cite{fortune2025hotrockssurveyiii}. Eclipse depth results are visualized in Figure \ref{fig:eclipse_depths}.}
\end{deluxetable*}

\subsection{LHS\,1478\,b}

For LHS\,1478\,b visit 2 we are unable to detect an eclipse, much like in the original Hot Rocks survey paper \citep{august_hot_2024}. This supports the conclusion from \cite{august_hot_2024} that the ``sinusoidal systematic" they observe during the predicted eclipse time is likely astrophysical in origin, as our FN-PCA detrending is (by design) unable to detect astrophysical signals and does not show signs of sinusoidal variation.

\subsection{TOI-1468\,b}

The results from \verb|Erebus| show that the first visit of TOI-1468\,b is shallower than the other two by $2\sigma$ (Figure \ref{fig:eclipse_depths}). The results shown in \cite{hotrocks2} also have this eclipse being shallower than the others. Our result for the two deep visits and the joint fit agrees with \cite{hotrocks2} within $1\sigma$, although for the first visit we find an eclipse depth that is lower by $\geq1\sigma$. In \cite{hotrocks2} they list a number of potential causes for why the first visit in particular was significantly shallower.

\subsection{LHS\,1140\,c}

For LHS\,1140\,c we get eclipse depths that are within $1\sigma$ of those found in \hbox{\cite{fortune2025hotrockssurveyiii}}. The mean and joint fit results of our exponential fit are closer to the ``GP aperture extraction" results reported in \hbox{\cite{fortune2025hotrockssurveyiii}} compared to our FN-PCA results. For those we get a shallower joint fit eclipse depth than found in the aforementioned paper, although still within $1\sigma$.

\newpage

\subsection{TRAPPIST-1\,b}

Our individual results are consistent with each other within about $1\sigma$. The mean of our FN-PCA individual fits agrees with the mean of the individual fits from \cite{greene_thermal_2023}, and our joint fit result is within error of their joint fit result. Our results differ by over $1\sigma$ for the individual fits of visits 2 and 3. Our FN-PCA and exponential fits are all within $1\sigma$ of each other. We provide a detailed plot for a single visit of TRAPPIST-1\,b (Figure \ref{fig:full plot}) which shows the FN-PCA detrended fit and the components which were detrended against.

\subsection{TRAPPIST-1\,c}

The results from \verb|Erebus| produce inconsistent eclipse depths between each visit, including a non-detection for visit 1
(Figure \ref{fig:eclipse_depths}). If we adjust the prior to only allow positive eclipse depths, we get a result in agreement with zero for this visit. If we do not force the eclipse depth to be positive for visits 2 and 3, our pipeline mistakes the eclipse egress for a large positive eclipse depth - the eclipse timing for this does not match the timing found in other visits, so we use the positive eclipse depth result. For the FN-PCA analysis of visit 4 we use a 2nd degree polynomial to model stellar variation in the observation that is not detected by the FN-PCA (Figure \ref{fig:eclipse_depths}, dashed error bars). Furthermore, our pipeline did not produce results which are in agreement with the values reported in \cite{zieba_no_2023} for the individual fits of visit 1 and visit 4. This is the case for both the systematic model and the exponential ramp, indicating that the discrepancy is not due to FN-PCA pipeline, but rather some other part of the data reduction process. We further investigated using a four pixel aperture closer in size to that used in \cite{zieba_no_2023} and find that we still see a non-detection for visit 1 and a very deep eclipse for visit 4. The joint fit eclipse depth for the 4 pixel aperture reduction is 337\,\tpm\,141 ppm, which is in agreement with our five pixel aperture result (312\,\tpm\,128, Table \ref{table:depths}).

\section{Telescope and Detector Systematics}\label{sec:systematics}

In order to investigate instrumental behaviour by examining overall trends in the eigenimage and eigenvalue time series produced by FN-PCA, we run the analysis on the published data shown in Section \ref{sec:data}, as well as all publicly available 15 $\mu$m eclipse data from the Hot Rocks survey \citep[GO 3730][]{hot-rocks-go} and a single 15 $\mu$m frame normalized observation of LP\,791-18\,d from GO\,6457. Both programs are currently on-going. From this, we group the FN-PCA eigenimages into 4 categories: ``detector-settling", ``centroiding", ``pixel-dominated", and ``random". We investigate the overall trend of the detector-settling component which is present in all of the data.

\subsection{``Detector-settling" component}\label{sec:detector-settling-res}

\begin{figure}[!t]
\centering
\includegraphics[width=\linewidth]{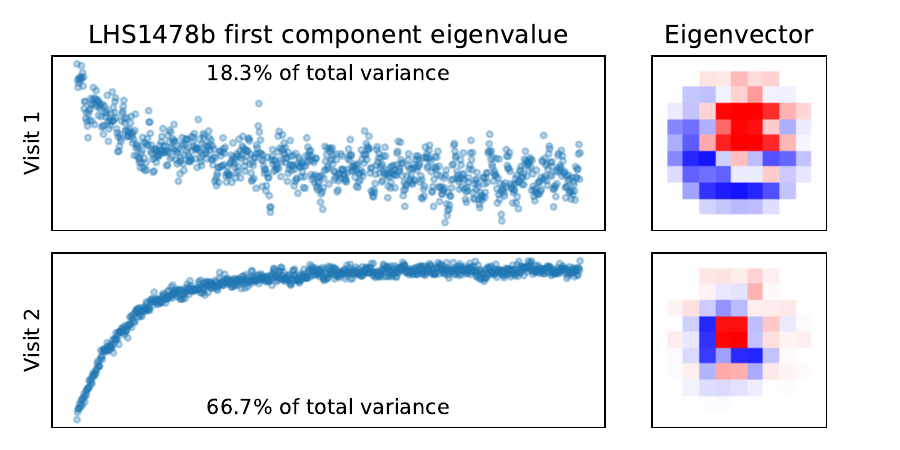}
\caption{The detector settling component either decays with a wide eigenimage covering most of the PSF (top) or ramps up with a more narrow eigenimage (bottom). The ramp-up component is less common and can have a lesser effect on the light curve depending on the size of the aperture and of the affected portion of the PSF due to it being narrower in general, despite being a stronger signal that explains more of the total variance found by PCA.}
\label{fig:settling slopes}
\end{figure}

All visits have a PC with an eigenimage showing pixel brightness decreasing in the center of the PSF and increasing in the outer portion of the PSF or vice-versa (Figure \ref{fig:settling slopes}). The eigenvalue is always an exponential ramp. 

We found that the detector-settling component can have either a positive or a negative slope (Figure \ref{fig:settling slopes}) which has similarly been observed in MIRI LRS \citep{bell2023_miri_lrs_settling_ramp, Dyrek_2024_miri_lrs_settling}. For most visits it is negative (the center of the PSF gets dimmer over the observation). We find that one visit each of TRAPPIST-1\,c, LHS\,1478\,b, and TOI-1468\,b, as well as two visits of LHS\,1140\,c, have a positively sloped detector-settling eigenvalue, where the affected portion of the PSF is much narrower than when the slope is negative (Figure \ref{fig:settling slopes}, right). For these positive-ramp detector-settling components the eigenvalue ramp is less noisy and the component explains a higher percentage of the total variance within the aperture compared to the negatively-sloped detector-settling components (Figure \ref{fig:settling slopes}, left). For the visits of TOI-1468\,b and LHS\,1478\,b showing this effect the affected portion of the PSF is entirely encircled by the aperture resulting in a lack of any exponential ramp at the start of the light curve which is seen in other visits (such as the exponential ramp seen for a visit of TRAPPIST-1\,b in Figure \ref{fig:full plot}). This effect also does not explain the aberrant visits of TOI-1468\,b and LHS\,1478\,b; the visits showing a shallow/undetected eclipse have the more common negatively-sloped detector-settling component. This is similarly the case for LHS\,1140\,c, where the eclipse depth is seemingly not correlated to the slope of the detector-settling effect.

In \cite{fortune2025hotrockssurveyiii} they find that the direction of the detector settling slope is related to the filter wavelength of the previous observation. In our analysis we have focused primarily on the length of the settling systematic as we find it to be a more detrimental effect when dealing with systematic noise, as depending on the length of the settling slope the eclipse signal itself can be affected.

\begin{figure*}[htp]
\centering
\includegraphics[width=0.9\linewidth]{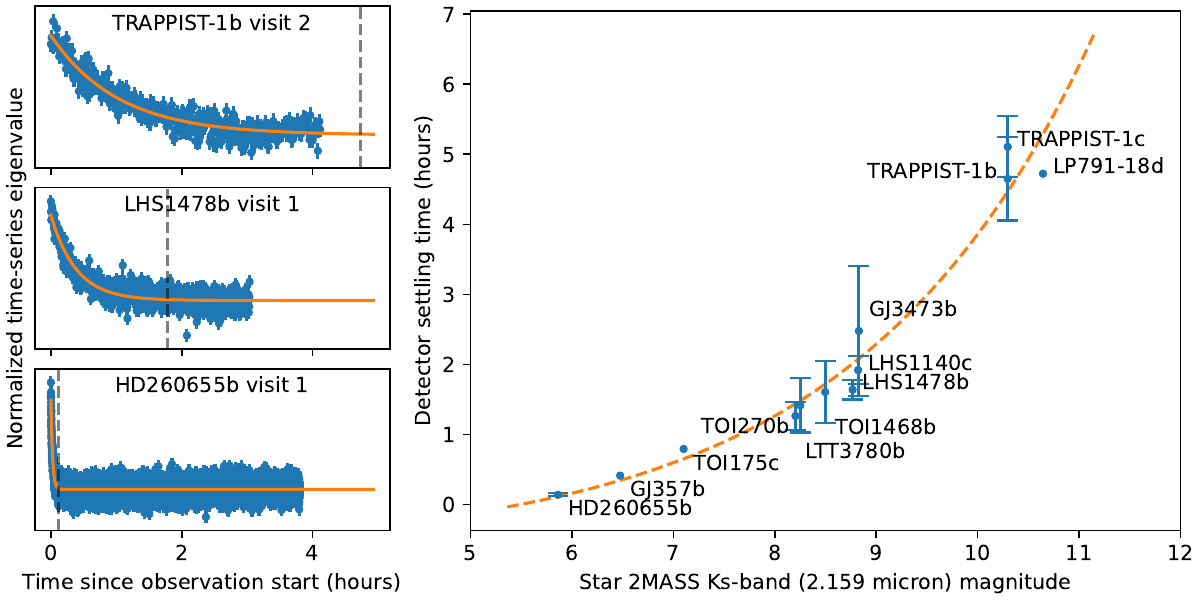}
\caption{(Left) Detector settling time per visit was taken as the time for 99\% of the detector-settling component eigenvalue (fit with an exponential) to decay. Three individual visits of planets orbiting high, medium, and low magnitude stars (TRAPPIST-1\,b, LHS\,1478\,b, and HD\,260655\,b respectively) are shown as examples of detector settling fitting. (Right) Best fit model between detector settling time and star Ks-band magnitude for all planets with available MIRI 15\,$\mu$m data. Each data point represents the average settling time across all available visits for that planet, with the error bars taken from the standard deviation of the settling time for all visits (for planets with only a single available visit the error is 0). These data points are generated from: four visits of GJ\,3473\,b; three visits of LHS\,1140\,c; two visits of LHS\,1478\,b; four visits of TOI\,270\,b; five visits of TRAPPIST-1\,b; three visits of TRAPPIST-1\,c (we discard one of the four available visits as an outlier); one visit of GJ\,357\,b; two visits of HD\,260655\,b; two visits of LTT\,3780\,b; three visits of TOI\,1468\,b; one visit of TOI\,175\,c; and one visit of LP\,791-18\,d. We label these components as representing detector settling due to their exponential shape and strong dependence on photon flux. For TRAPPIST-1\,b (top-left) the settling time is greater than the duration of the observation, for LHS\,1478\,b (middle-left) the settling time is greater than the midpoint of the observation, and for HD\,260655\,b (bottom-left) the settling time is near immediate. The best-fit for the detector settling time relationship is: $T_{settling} = 0.063e^{0.427m_K} -0.657$ hours.}
\label{fig:settling}
\end{figure*}

We find that the length of the detector-settling eigenvalue is directly related to the magnitude of the star. We modeled the settling ramps with an exponential decay for each visit which was fit using \verb|scipy.optimize.curve_fit| (Figure \ref{fig:settling}, left). The timescale over-which the exponential ramp stabilizes (99\% of the signal decayed) is correlated to the magnitude of the star with an exponential curve (Figure \ref{fig:settling}, right) given by the equation

\begin{figure}[!t]
\centering
\includegraphics[width=\linewidth]{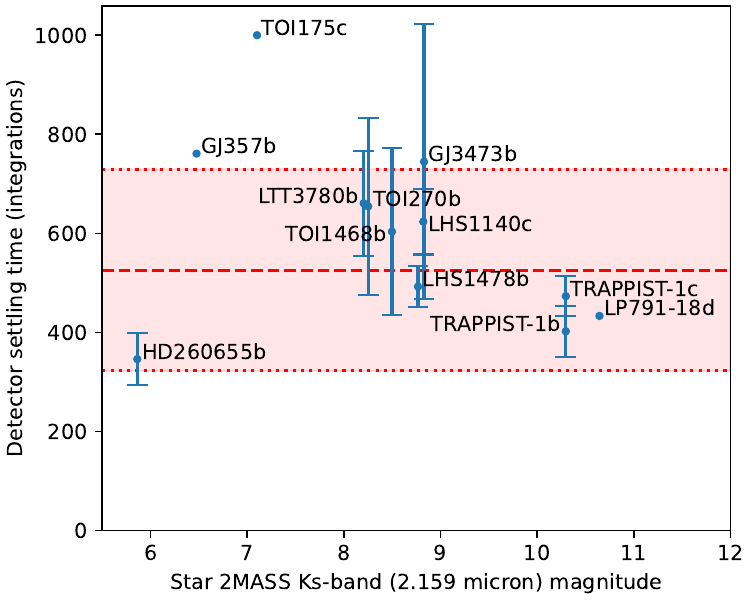}
\caption{The number of integrations for the detector to settle as a function of the star's Ks band magnitude for all planets with available MIRI 15 $\mu$m data. 525\,\tpm\,203 integrations are required for the detector to settle, which is marked as the red region. TOI175\,c appears as a notable outlier, however this data point is based solely on a single available visit and does not take into account possible flux contributions by the two other planets in its star system \citep{toi175c}.}
\label{fig:settling_integrations}
\end{figure}

\begin{equation}
    T_{settling} = 0.063e^{0.427m_K} -0.657 \text{ hours.}
\end{equation}

Due to this strong dependence between photon flux and ramp settling time, we conclude that this first component represents detector settling. For the dimmest stars observed the detector settling time was longer than the observation length. For the brightest stars observed the settling component is replaced by the centroiding components along both diagonal axes as the most prominent sources of systematic noise. We find that 525 \tpm 203 integrations are required for the detector to fully settle (Figure \ref{fig:settling_integrations}).

\FloatBarrier

\subsection{``Centroiding" component}\label{sec:centroiding-res}

\begin{figure}[h!]
\centering
\includegraphics[width=0.4\linewidth]{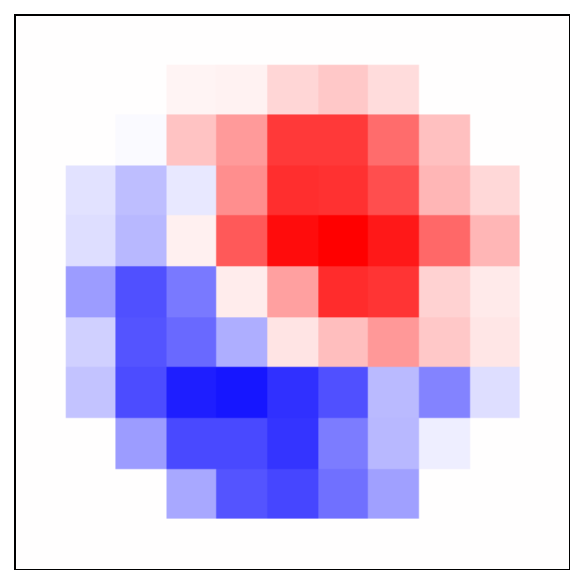}
\includegraphics[width=0.4\linewidth]{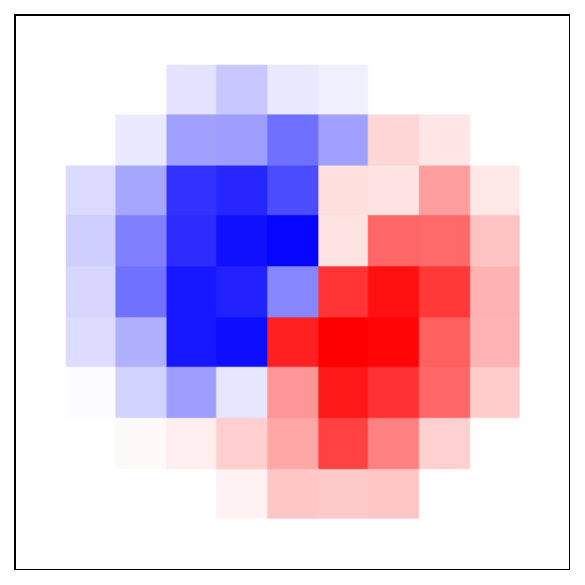}
\caption{The first and fourth principal components of TOI-1468\,b visit 1 show the Gaussian centroid moving along two perpendicular axes. The positive diagonal explains 13.9\% of the total variance and the negative diagonal explains only 3.6\%.}
\label{fig:HD260655b centroiding}
\end{figure}

\FloatBarrier

The Gaussian centroid of the PSF moves throughout the visit due to movement of the JWST pointing. This can manifest in two components, each representing movement along a perpendicular axes (Figure \ref{fig:HD260655b centroiding}).

In the majority of visits we investigated, the positive-diagonal axis is the 2nd highest ranked component. For brighter stars the positive-diagonal centroiding component represents the most variance. The 2nd perpendicular centroiding component (negative-diagonal) is always ranked below the positive-diagonal component; in the dimmest star analyzed (TRAPPIST-1) it does not appear in the top 5 highest ranked components, while in the brightest star analyzed (HD260655b) it appears as the 2nd strongest component in all visits. The rank of these components increases as the brightness of the star increases.

\subsection{``Pixel-dominated" components}\label{sec:pixel-dominated-res}

\begin{figure}[ht]
\centering
\includegraphics[width=0.4\linewidth]{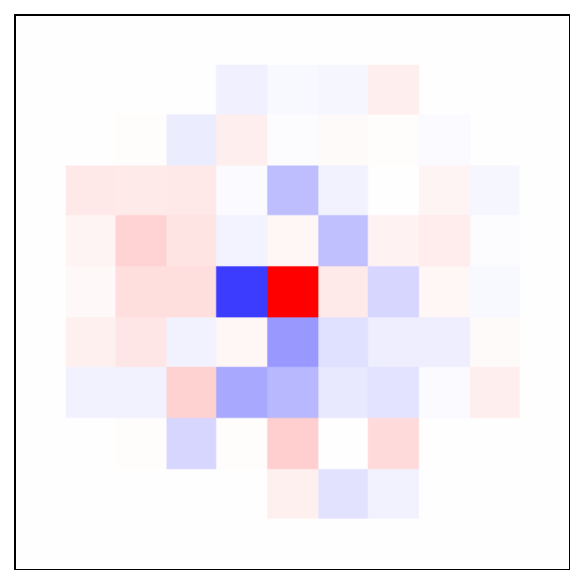}
\includegraphics[width=0.4\linewidth]{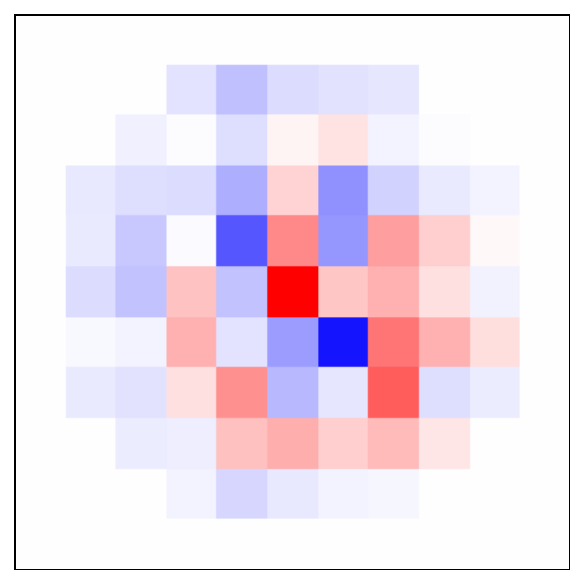}
\caption{Pixel-dominated principal components from visits of TRAPPIST-1\,c (left) and TOI-1468\,b (right).}
\label{fig:pixel-dominated}
\end{figure}

The majority of the remaining components for the dimmer stars (K-band magnitude 9 to 12) are pixel-dominated where the eigenimage shows a single pixel being correlated to either another pixel or a group of pixels (Figure \ref{fig:pixel-dominated}). These pixel effects may be related to the ``Brighter-Fatter Effect" which has been observed in the MIRI detector pixels \citep{Argyriou_2023}.

\subsection{``Random" components}\label{sec-random-res}

\begin{figure}[ht]
\centering
\includegraphics[width=0.4\linewidth]{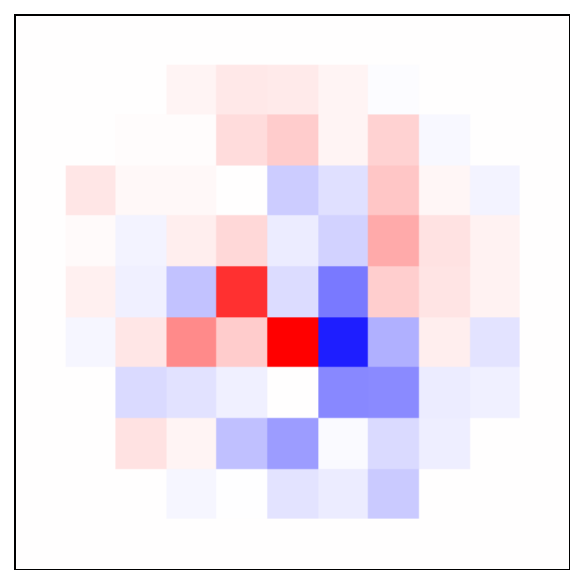}
\includegraphics[width=0.4\linewidth]{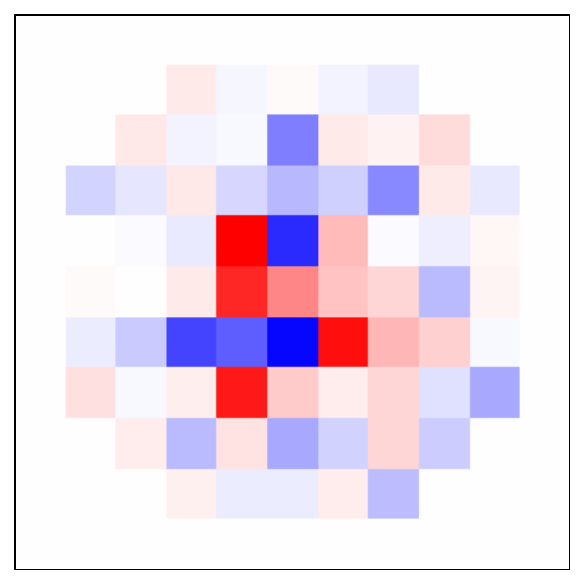}
\caption{Random principal components from two visits of TRAPPIST-1\,b.}
\label{fig:random}
\end{figure}

This category makes up the rest of the components, which are difficult to categorize by eye. These are more common for the brighter stars (K-band magnitude 5 to 9). Examples are shown in Figure \ref{fig:random}.

\FloatBarrier

\section{Atmosphere and surface analysis}\label{sec:scientific-results}

We simulated a number of emission spectra for TRAPPIST-1\,b, TRAPPIST-1\,c, TOI-1468\,b, LHS\,1478\,b, and LHS\,1140\,c, assuming various bare rock and atmospheric compositions (Figure \ref{fig:surface_models}). The bare rock models were chosen to represent a range of potential surface albedos, while the atmosphere models were largely designed to probe the CO$_{2}$ absorption band at 15 $\mu$m. Here we compare them to potential atmospheric and surface models for these planets.

\subsection{Atmospheric models with \texttt{SCARLET}} \label{subsec:atm}

The \verb|SCARLET| modeling framework was employed to generate 1D self consistent atmospheric models \citep{scarlet_1, scarlet_2, scarlet_3, 2019ApJ...887L..14B, 2019NatAs...3..813B, benneke2024jwstrevealsch4co2, 2021AJ....162...73P, Pelletier_Benneke_Chachan_Bazinet_Allart_Hoeijmakers_Lavail_Prinoth_Coulombe_Lothringer_et_al._2024, Roy_Benneke_Piaulet_Crossfield_Kreidberg_Dragomir_Deming_Werner_Parmentier_Christiansen_et_al._2022, Roy_2023, Piaulet_Ghorayeb_2024, Bazinet_2024, Monaghan_Roy_Benneke_Crossfield_Coulombe_PiauletGhorayeb_Kreidberg_Dressing_Kane_Dragomir_etal._2025}. These models assume a nongray, radiative, convective temperature profile and a well-mixed composition of 100ppm of CO$_{2}$ within an O$_{2}$ or H$_{2}$O dominated atmosphere. The models solve for hydrostatic equilibrium and radiative transfer iteratively to produce a theoretical temperature-pressure profile of the planet. Once the model converges to stability, the associated emission spectrum from the secondary eclipse is computed. No tidal heating is assumed in these models.

\subsection{Surface models with \texttt{JESTER}} \label{subsec:jester}

The bare rock models were calculated using \verb|JESTER|, which simulates the surface emissions of a bare rock exoplanet given a surface composition and stellar model (Monaghan et al. (in prep.)). The wavelength-dependent hemispherical reflectance of lab samples measured in \citet{Paragas_Knutson_Hu_Ehlmann_Alemanno_Helbert_Maturilli_Zhang_Iyer_Rossman_2025} were used to approximate two ultramafic surfaces of varying grain size on each of the five planets, alongside a dark bare rock with zero reflectivity. These values were converted into the corresponding wavelength-dependent emissivity and directional hemispherical reflectance following Hapke theory in order to model the planet's reflected and emitted surface flux \citep{Hapke_2012}. We account for the nonuniform temperature gradient on the planet's dayside for all bare-rock models by separating the hemisphere into a number of circular regions at an angle $\theta$ away from the substellar point. The spectral flux density of each region is then calculated such that the sum of the reflected and thermal emission from each region is equal to the incident net flux from the host star onto the annular section \citep[][]{Monaghan_Roy_Benneke_Crossfield_Coulombe_PiauletGhorayeb_Kreidberg_Dressing_Kane_Dragomir_etal._2025}. The total spectral flux density of the planet's dayside is then calculated using the sum of the reflected and emitted flux density, weighted by the area of each annular section.

\subsection{Treatment of the stellar flux} \label{subsec:stellarflux}

The models from both \verb|SCARLET| and \verb|JESTER| require the use of a stellar model in order to appropriately calculate the planet's outgoing flux. For consistency, a SPHINX stellar model is used to simulate the host star's flux for each planet, using an assumed C/O ratio of 0.7. \citep{Iyer_Line_Muirhead_Fortney_Gharib-Nezhad_2023, 2024ESS.....563103I}. The model stellar flux is then used in \verb|JESTER| to calculate the temperature gradient of the planet and estimate the surface flux density of the planet's dayside. However, when modeling the expected eclipse depth in Figure \ref{fig:surface_models}, the model stellar flux is scaled using the absolute calibrated flux from host star received by Earth in the F1500W filter for each system, measured to be 2.528 mJy for TRAPPIST-1 \citep{Ducrot_2024}, 10.54 mJy for TOI-1468 \citep{hotrocks2}, and 9.177 mJy for LHS\,1140 \citet{fortune2025hotrockssurveyiii}. The expected fluxes of the SPHINX stellar models for each system weighted by the F1500W filter transmission is calculated to be 2.665 mJy for TRAPPIST-1, 11.727 mJy for TOI-1468, and 8.727 mJy for LHS\,1140. The stellar flux used in the eclipse depth calculation is scaled such that the model flux matches the absolute calibrated flux of the host star as measured by MIRI. Note that we do not perform flux recalibration of the models for LHS\,1478\,b as the SPHINX models were found to suitably model the host star's flux \citep{august_hot_2024}.

\subsection{Results}\label{subsec:results}

We recalculate the dayside brightness temperature $T_{d}$ of each planet using the FN-PCA measured eclipse depths using a simple minimization function. The photon flux $F_{\gamma}$ measured in the F1500W filter from the planet (assuming it emits as a uniform temperature blackbody) and host star is calculated as:

\begin{equation}
    F_{\gamma,p} = \int \frac{\pi B_{p}(T_{d}) \lambda}{hc} W_{\lambda}\,d\lambda
\end{equation} \label{eq:photonfluxplanet}
\begin{equation}
    F_{\gamma,star} = \int \frac{F_{star} \lambda}{hc}W_{\lambda}\, d\lambda
\end{equation}\label{eq:photonstarplanet}

\noindent where $B_{p}(T_{d})$ represents the Planck function, $W_{\lambda}$ the throughput of the F1500W filter, and $F_{star}$ the stellar flux density of the rescaled SPHINX model. For each planet, we find the value $T_{d}$ such that $\frac{R_{p}^{2}}{R_{star}^{2}}\frac{F_{\gamma,p}(T_{d})}{F_{\gamma,star}}$ is equal to the FN-PCA measured eclipse depth.

\begin{figure*}[t!]
    \centering
    \includegraphics[width=1\linewidth]{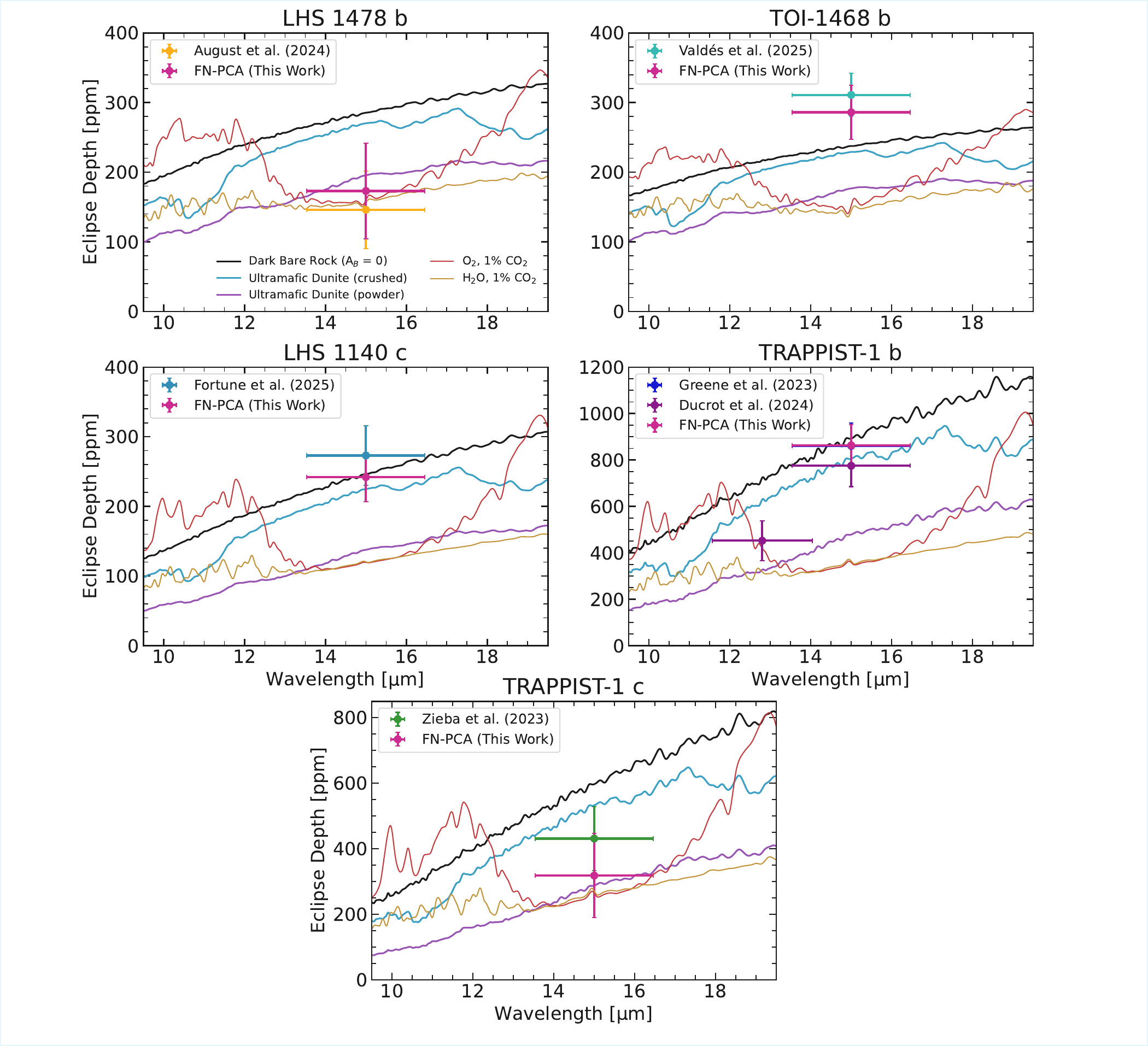}
    \caption{Simulated emission spectra for LHS\,1478\,b, TOI-1468\,b, LHS\,1140\,c, TRAPPIST-1\,b, and TRAPPIST-1\,c compared to the measured eclipse depths from our FN-PCA calculation, alongside the previous literature results reported in Table \ref{table:depths} (in addition to the values calculated by \citet{Ducrot_2024} for TRAPPIST-1\,b). The joint fit results are plotted for all planets except for LHS\,1478\,b, which only plots the measured eclipse depth from the first visit. The Ultramafic Dunite spectra were modeled using the dunite xenolith wavelength-dependent spherical reflectance from \citet{Paragas_Knutson_Hu_Ehlmann_Alemanno_Helbert_Maturilli_Zhang_Iyer_Rossman_2025}.}
    \label{fig:surface_models}
\end{figure*}

For LHS\,1478\,b, we compare our surface and atmospheric models to the result of visit 1 due to the inconsistency between visits 1 and 2 and the joint fit detecting no eclipse (Figure \ref{fig:eclipse_depths}). From the first visit, we measure a brightness temperature of $583^{+108}_{-115}$K, in agreement with the temperature reported by \citet{august_hot_2024} of $491\pm102$K. We rule out a dark bare rock composition at $\sim2\sigma$, and find that the measured eclipse depth is consistent with an atmosphere enriched with CO$_{2}$, similarly to what is concluded in \citet{august_hot_2024}. For a uniform temperature bare rock with zero heat redistribution ($f = \frac{2}{3}$), the FN-PCA brightness temperature is most consistent with a surface bond albedo $A_{B}$ of 0.66, as calculated using:

\begin{equation}
    T_{d} = T_{star}\sqrt{\frac{R_{star}}{a_{p}}}\big(1-A_{B}\big)^{\frac{1}{4}}f^{\frac{1}{4}}.
\end{equation}\label{eq:barerock}

Because the second observation had to be discarded, we make no strong conclusions regarding the planet's composition and await the LHS\,1478\,b follow up observations in the forthcoming GO\,7675 program.

For TOI-1468\,b, we find that the planet is hotter than expected for a bare rock (Figure \ref{fig:surface_models}) as is found in \cite{hotrocks2}. We find a dayside brightness temperature of $964^{+78}_{-83}$K, which agrees within $1\sigma$ of $1024 \pm 78$K reported by \cite{hotrocks2}. A number of astrophysical and systematic hypotheses are presented by \citet{hotrocks2} to explain the origin of the measured dayside temperature, including a thermal inversion in the planet's atmosphere \citep{thermal_inversion_2011ApJ...742L..19M, thermal_inversion_2013cctp.book.....M, thermal_inversion_Castan_2011, thermal_inversion_Zilinskas_2022}, induction heating from the host star's magnetic field \citep{induction_heating_2014MNRAS.441.2361V, induction_heating_2016PhR...663....1S, induction_heating_2020MNRAS.498.5684D, induction_heating_Kislyakova_2018}, and unaccounted instrumental artifacts in MIRI \citep{Dyrek_2024_miri_lrs_settling, miri_artifacts_libralato2024highprecisionastrometryphotometryjwstmiri, miri_artifacts_Morrison_2023}. For the latter, our FN-PCA analysis does not reveal any MIRI systematics that could have deepened the eclipse depth. Future observations are necessary in order to further understand the nature of the deep eclipse.

For LHS\,1140\,c we measure a dayside brightness temperature of $528^{+32}_{-36}$K, which is lower than the value reported by \citep{fortune2025hotrockssurveyiii} of $561\pm44$ by $\sim1\sigma$. However, our results further rule out an atmospheric presence on the planet, as the FN-PCA eclipse depth is largely consistent with a low-albedo bare rock composition.

For TRAPPIST-1\,b we find a dayside brightness temperature of $499\pm24$K, consistent with both brightness temperatures reported by \citet{greene_thermal_2023} and \citet{Ducrot_2024} of $503^{+26}_{-27}$K and $478\pm27$K, respectively. Our results further support TRAPPIST-1\,b's composition as an atmosphereless bare rock with a low surface albedo \citep{ih_constraining_2023, Ducrot_Lagage_Min_Gillon_Bell_Tremblin_Greene_Dyrek_Bouwman_Waters_etal._2025}. 

For TRAPPIST-1\,c we find a dayside brightness temperature of $344^{+43}_{-52}$K. This temperature is lower than the value previously reported in \cite{zieba_no_2023} of $380 \pm 31$K by $\sim1\sigma$, who found that the eclipse depth of the planet was most consistent with a semi-reflective bare rock composition or a thin atmosphere of O$_{2}$ and CO$_{2}$ \citep{Lincowski_Virtual_Planetary_Laboratory_Meadows_Zieba_Kreidberg_Morley_Gillon_Selsis_Agol_Bolmont_et_al._2024}. Notably, the FN-PCA eclipse depth further supports the atmospheric scenario for TRAPPIST-1\,c, with an eclipse depth consistent with both CO$_{2}$ models simulated in Figure \ref{fig:surface_models}. Low eclipse depths may also be associated with high albedo clouds within the planet's atmosphere \citep{Mansfield_Kite_Hu_Koll_Malik_Bean_Kempton_2019}. Alternatively, highly reflective bare rock surfaces may also be consistent with the measured eclipse depth. Ultramafic surfaces formed by the partial melting of rock are of particular note for planets orbiting M dwarf stars, as the high albedo may lead to false positives for atmospheric detection \citep{Mansfield_Kite_Hu_Koll_Malik_Bean_Kempton_2019, Hammond_Guimond_Lichtenberg_Nicholls_Fisher_Luque_Meier_Taylor_Changeat_Dang_etal._2025, Paragas_Knutson_Hu_Ehlmann_Alemanno_Helbert_Maturilli_Zhang_Iyer_Rossman_2025}.

There are a number of caveats that may affect the accuracy of the simulated emission spectra. Most notably, imperfect knowledge of the stellar spectrum used to simulate the planet's emission spectra may lead to misleading interpretations of the planet's composition \citep[e.g.,][Monaghan et al. (in prep.)]{fauchez2025stellarmodelslimitexoplanet}. Although scaling the stellar model using the absolute flux calibration as outlined in Section \ref{subsec:stellarflux} should result in a more accurate interpretation of the eclipse depth within the MIRI F1500W bandpass, it may lead to inaccuracies outside of the filter's throughput as we lack knowledge of the host star's bolometric spectrum. \citet{fauchez2025stellarmodelslimitexoplanet} noted that for TRAPPIST-1, although SPHINX models overestimated the star's mid-IR flux, they were an adequate fit to the star's flux near-IR emissions below 5$\mu$m, where the majority of the star's flux is sourced from. It is for this reason that \verb|JESTER| uses the uncalibrated SPHINX models to simulate the planet's dayside flux density, as scaling the entirety of the host star's spectrum beforehand would result in an improperly scaled temperature gradient on the planet's dayside.

Model accuracy of the simulated atmospheres may be impacted by the presence of clouds, winds, and photochemistry on any of the planets. The mixing ratios of the chosen molecules are also unlikely to be uniform across the planet's altitude, and other elemental and molecular species unaccounted for in our models may further impact the measured infrared emissions.

For bare rock planets, chemical contaminants, mineralogical mixtures, and space weathering are all known to affect the reflectivity and emissivity of the planet's surface \citep[e.g.,][]{Paragas_Knutson_Hu_Ehlmann_Alemanno_Helbert_Maturilli_Zhang_Iyer_Rossman_2025,    2015aste.book..597B, 2019A&A...627A..43D, Lyu_Koll_Cowan_Hu_Kreidberg_Rose_2024, Mansfield_Kite_Hu_Koll_Malik_Bean_Kempton_2019}. Furthermore, the emissivity of different mineralogical compositions is largely dependent on temperature \citep[e.g.,][]{2013E&PSL.371..252H, 2020E&PSL.53416089F,  2021PSJ.....2...43T, 2021Icar..35414040P}. Thus, the laboratory spherical reflectance of each composition may not accurately describe the conditions on each planet, which vary in surface temperature. In addition, the surface models provided by \citet{Paragas_Knutson_Hu_Ehlmann_Alemanno_Helbert_Maturilli_Zhang_Iyer_Rossman_2025} only provide spherical reflectance for wavelengths shorter than $20\mu$m. Although the contribution of flux from the star above 20$\mu$m is small, completely ignoring these wavelengths would result in an artificially inflated dayside temperature (Monaghan et al. (in prep.)). Thus, in order to more accurately account for the energy balance of a true bare rock planet across the entire bolometric wavelength regime, a uniform reflectivity is assumed for the ultramafic surfaces at $\lambda > 20\mu$m of $\sim0.35$ and $\sim0.5$ for the crushed and powdered surfaces respectively. This modification avoids the overestimation of the planet's dayside temperature and produces adequate eclipse depth models at $\lambda < 20\mu$m.

\FloatBarrier

\section{Conclusion}\label{sec:conclusion}

We introduced a new lightcurve detrending method, Frame-Normalized Principal Component Analysis (FN-PCA) and showed that it is a viable systematic model for fitting against pixel-level noise in the MIRI $15\micron$ data. We presented a new data reduction pipeline (\verb!Erebus!) which uses this approach to data analysis, which is now ready to be used on the results of the 500 hour Rocky Worlds DDT survey as the data is released. This pipeline is open source at \hyperlink{https://github.com/nicholasconnors/erebus}{https://github.com/nicholasconnors/erebus} and easily adaptable to analyze MIRI photometry data in filters other than F1500W.

We reanalyzed results from LHS\,1478\,b \citep{august_hot_2024}, TOI-1468\,b \citep{hotrocks2}, LHS\,1140\,c \citep{fortune2025hotrockssurveyiii}, TRAPPIST-1\,b \citep{greene_thermal_2023}, and TRAPPIST-1\,c \citep{zieba_no_2023} using FN-PCA detrending with our \verb!Erebus! pipeline and compared them to different planet surface and atmospheric compositions (Figure \ref{fig:surface_models}). Using this FN-PCA technique, we provide a deeper insight into the nature of the systematics in the JWST MIRI photometric time series. Furthermore, it provides the advantage that no parameterized form needs to be imposed. We crosscheck our new technique against results previously reported from JWST MIRI data sets, and find that our results are largely in agreement with the aforementioned publications which used conventional parameterized detrending techniques, with the exception of two individual fits each of TRAPPIST-1\,c and TRAPPIST-1\,b (with the joint fits agreeing within $1\sigma$).

Further looking across multiple data sets, we inspected the eigenvalues for the principal components corresponding to detector settling and calculate a relationship between the detector settling time and target star's K-band magnitude (Figure \ref{fig:settling}). For the brightest stars (less than 7 magnitude) the settling time is on the order of minutes, while for the dimmest starts (greater than 11 magnitude) it is on the order of several hours. This relationship can be used to estimate how long a JWST observation must be for the detector-settling to stop being an important source of systematic noise based on the star's K-band magnitude. This relationship can be used to inform the planning of MIRI F1500W observations for rocky exoplanets in studies like the 500 hour Rocky Worlds DDT.

\section*{Acknowledgments}

We thank the anonymous referee for their thorough review and helpful suggestions which improved this work. All of the data presented in this paper were obtained from the Mikulski Archive for Space Telescopes (MAST) at the Space Telescope Science Institute using the following DOI: \href{https://dx.doi.org/10.17909/h2e9-0j87}{10.17909/h2e9-0j87}. N.C. and  C.M. acknowledge financial support from the
University of Montreal. C.M. further acknowledges
financial support from Jean-Marc Lauzon and from
the Natural Sciences and Engineering Research Council
(NSERC) of Canada. This work
was made with the support of the Institut Trottier de
Recherche sur les Exoplanetes (iREx).

We used the following code resources in our data analysis: \verb!astropy! \citep{astropy:2013, astropy:2018, astropy:2022}, \verb!emcee! \citep{Foreman_Mackey_2013}, \verb!NumPy! \citep{harris2020array_numpy}, \verb!Eureka!! \citep{Bell2022}, \verb!batman! \citep{batman}, \verb!sklearn! \citep{scikit-learn}, \verb!matplotlib! \citep{matplotlib}, \verb!jwst! \citep{jwst_pipeline}, \verb!h5py! \citep{collette_python_hdf5_2014}, \verb!SciPy! \citep{2020SciPy-NMeth}, \verb!uncertainties!, \verb!pydantic!, and \verb!corner! \citep{corner}.

\bibliography{references}



\end{document}